# GPS Information and Rate Tolerance

—Clarifying Relationship between Rate Distortion and Complexity Distortion

Chenguang Lu*, Independent Researcher
Ma Anshan, P. R. China
Survival99@gmail.com

*Abstract*—I proposed rate tolerance and discussed its relation to rate distortion in my book "A Generalized Information Theory" published in 1993. Recently, I examined the structure function and the complexity distortion based on Kolmogorov's complexity theory. It is my understanding now that complexity-distortion is only a special case of rate tolerance while constraint sets change from fuzzy sets into clear sets that look like balls with the same radius. It is not true that the complexity distortion is generally equivalent to rate distortion as claimed by the researchers of complexity theory. I conclude that a rate distortion function can only be equivalent to a rate tolerance function and both of them can be described by a generalized mutual information formula where $P(Y|X)$ is equal to $P(Y|\text{Tolerance})$. The paper uses GPS as an example to derive generalized information formulae and proves the above conclusions using mathematical analyses and a coding example. The similarity between the formula for measuring GPS information and the formula for rate distortion function can deepen our understanding the generalized information measure.

## I. Introduction

There are two reasons for me to write this paper. Firstly, in my book "*A Generalized Information Theory*" published in 1993 [1], I introduced how Shannon Information theory [2] can be generalized for more general fields including the fields of linguistic and sensory communications and proposed two functions: rate tolerance (or rate-of-limiting-errors) $R(A_J)$[1] and rate-of-keeping-precision $R(G)$ as new visions of Shannon's rate distortion. Later I published some papers for further discussions [3] [4]. Recently, I examined the structure function [5] and complexity distortion [6] based on Kolmogorov's complexity theory. I found that complexity distortion is only a special case of rate tolerance; general equivalence between complexity distortion and rate distortion claimed by the researchers of complexity theory does not exist. Secondly, I found that GPS was a good model for explaining generalized information measure and could help us to understand rate distortion theory better. In this paper, I introduce the generalized information formulae first, and then discuss the equivalences between those functions related to data compression.

---
\* Author's homepage: http://survivor99.com/lcg/english
[1] In this paper, I use $R(T)$ instead of $R(A_J)$.

## II. GPS Information: From Statistical Information to Predictive Information

### A. GPS Accuracy and Set-Bayesian Formula

Global Positioning System (GPS) is a commercial and widely used global positioning system. We assume that the possible actual positions of an object surrounding the pointed position by a GPS are normally distributed.

Assume that the information source is the position of a stolen car with a GPS tracker. For simplicity, we use one dimension discrete random variables $X \in A=\{x_1, x_2 \ldots\}$ and $Y \in B=\{y_1, y_2 \ldots\}$ to denote the car's actual position and the displayed position respectively, and $y_j$ is $\hat{x}_j$ or prediction "$X$ is about $x_j$". If $X$ before the prediction is equiprobable, then

$$P(x_i \mid y_j) = \frac{1}{\sqrt{2\pi}\sigma} \exp\left[-\frac{(x_i - x_j)^2}{2\sigma^2}\right] \qquad (1)$$

See Fig.1 for the illustration of (1) and (2).

A common notation of GPS accuracy is root mean square error (RMS)[2], in which DRMS=10m means σ=10 meters and $X$ locates in the 10 meter circle surrounding $x_j$ with 68.2% possibility; 2DRMS=10m means 2σ=10 meters and $X$ locates in the same circle with 95.44% possibility. Another common notation of GPS accuracy is circular error probable (CEP) where CEP=10m means that $X$ locates in a 10 meter circle surrounding $x_j$ with 50% possibility.

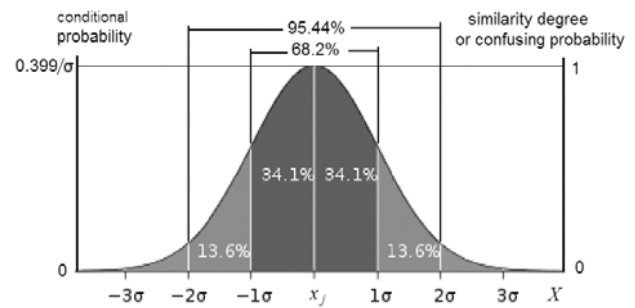

Figure 1. Tolerance function of GPS reflects confusing probability or similarity degree of $X$ with $x_j$

---
[2] http://www.igage.com/mp/GPSAccuracy.htm

Note that generally $P(X)$ is not equiprobable and $P(X|y_j)$ is not in normal distribution. Actually, the distribution function for GPS accuracy is independent of information source. It is better written as

$$c(x_i, x_j) = \exp\left[-\frac{(x_i - x_j)^2}{2\sigma^2}\right] \quad (2)$$

We call $c(x_i, x_j)$ as confusing probability or similarity degree of $x_i$ with $x_j$. The maximum of $c(x_i, x_j)$ is 1, yet the sum of conditional probability $P(X|y_j)$ is 1.

The confusing probability may be obtained from the statistics of a random set [7] [8]. Assume that we do $N$ times of experiments for $x_j$. Each time we determine a subset of $A$ where all $x_i$ in the subset are confused with $x_j$. If $N_i$ subsets includes $x_i$, then

$$c(x_i, x_j) = \lim_{N \to \infty} N_i / N \quad (3)$$

Let us define fuzzy set [9] $A_j=\{$all $x_i$ confused with $x_j \}$, and let $Q(A_j|x_i)$ denote the membership grade of $x_i$ in $A_j$, and $Q(A_j|x_i)= c(x_i, x_j)$. $Q(A_j|x_i)$ can also be considered as the similarity degree of $x_i$ with $y_j$ or the posterior logical probability of $y_j$="$X$ is about $x_j$". The prior logical probability is its average:

$$Q(A_j) = \sum_i P(x_i)Q(A_j | x_i) \quad (4)$$

This is just the probability of a fuzzy event proposed by Zedah [10]. If $x_i \in A_j$, to get the conditional probability of $x_i$, we have set-Bayesian formula [4] or generalized Bayesian formula [8]:

$$P(x_i | A_j) = P(x_i | x_i \in A_j) = P(x_i)Q(A_j | x_i)/Q(A_j) \quad (5)$$

In discussing rate distortion function and rate tolerance function bellow, we shall see the above formulae about $Q(A_j)$ and $P(x_i|A_j)$ frequently. If $A_j$ is a clear set, the illustration of $P(X|A_j)$ is shown in Fig. 3 bellow.

### B. From Statistical Information to Predictive Information

To measure information amount from a single event such as a GPS reading, it is not feasible to use Shannon entropy $H(X)$ or mutual information $I(X;Y)$ because they are only for average information [2]. So, some researchers suggest to use the logarithmic part of Shannon's mutual information formula to measure information from a single message $y_j$ about $x_i$. The formula is [11]

$$I(x_i; y_j) = \log \frac{P(x_i | y_j)}{P(x_i)} \quad (6)$$

However, (6) will result in negative information amount as $P(x_i|y_j)<(P(x_i)$ so that it is not widely accepted.

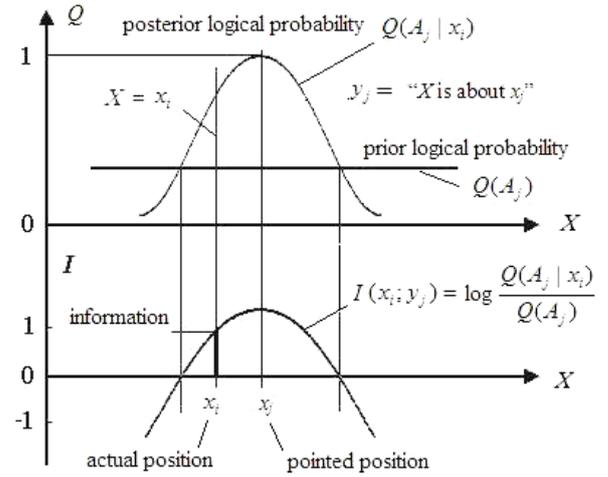

Figure 2. Illustrating predictive information formula

I accept this formula. Since information reflects saved coding length, naturally, negative information reflects increased coding length due to a wrong forecast or a lie.

Moreover, to measure GPS information, we also need factual test because according to common sense, if positioning is correct, the information amount is bigger; if positioning is inaccurate or wrong, the information amount is less or even negative. So, I call GPS information as predictive information and Shannon information as statistical information.

Replacing "$y_j$" in (6) with "$y_j$ is true" or "$x_i \in A_j$", we have predictive information formula:

$$I^*(x_i; y_j) = \log \frac{P(x_i | y_j \text{ is true})}{P(x_i)} \quad (7)$$
$$= \log \frac{P(x_i | A_j)}{P(x_i)} = \log \frac{Q(A_j | x_i)}{Q(A_j)}$$

Note that beside $P(X)$ and $Q(A_j|X)$, we also need the fact $X=x_i$.

It can been seen from Fig. 2 that the closer $x_i$ is to $x_j$, the bigger the information amount is; with the deviation increasing, the information amount will decrease or become negative; the less the $Q(A_j)$ is, the bigger the absolute value of information amount is. Two factors determine the quantity of $Q(A_j)$. One is GPS accuracy, i.e. $\sigma$. Another is $P(X)$ covered by $Q(A|X)$. So we can conclude that the more accurate and more unexpected the prediction is, the bigger the absolute value of information amount is. Meanwhile the mistake is more possible and the test is more severe for small $Q(A_j)$. If $Q(A_j|x_i)$ is big enough, which means that the prediction can withstand the test, the information amount will be bigger. This formula corresponds to Popper's information criterion for the advancement of scientific theories (in [12], section 10-II) very much.

### C. Information of An Ideal GPS or A Universally Necessary Statement

An ideal GPS is defined as

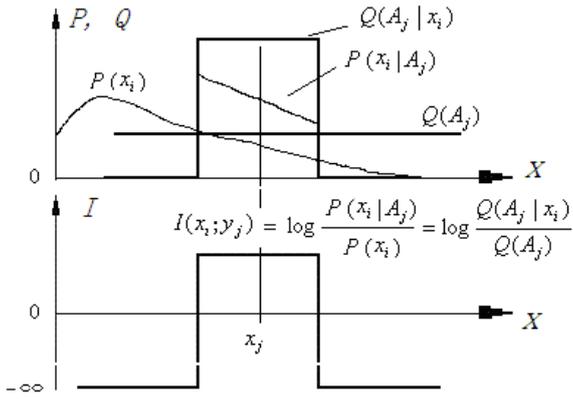

Figure 3.  Information conveyed by an ideal GPS or a universally necessary statement

1) the confusing probability only takes value from {0,1};

2) it is claimed that the object must locate in a circle with 100% possibility.

For an ideal GPS, the information formula becomes

$$I^*(x_i; y_j) = \log \frac{Q(A_j | x_i)}{Q(A_j)} = \begin{cases} -\log Q(A_j), & x_i \in A_j \\ -\infty, & x_i \notin A_j \end{cases} \quad (8)$$

Its illustration is shown in Fig. 3. It is easy to prove that information will be $-\infty$ if there is one $x_i \notin A_j$. This means that when we code a source with probability distribution $P(X|A_j)$, if there is one $x_i \notin A_j$, coding must fail no matter how long the codes are.

According to Popper's falsification theory, one opposite example is enough to falsify a scientific statement [12]. The above formula just reflects this Popper's thought. For necessarily true and necessarily false statements, there is always $Q(A_j|x_i)/Q(A_j)=1$ and $I(x_i; y_j)=0$. This conclusion also corresponds to Popper's thought.

In daily language, since information receivers always assume that unexpected event is possible and understand statements in fuzzy mode, negative infinite information rarely happens.

### D. Generalized Kullback Formula and Its Applications to Forecasts

How do we measure the average information of $y_j$ about $X$? In earlier days, I used Kullback formula:

$$I_k(X; y_j) = \sum_i P(x_i | A_j) \log \frac{P(x_i | A_j)}{P(x_i)} \quad (9)$$

where $P(X|Y)=P(X|Y$ is true$)$ is assumed, i.e. for all $x_i$, $P(x_i | y_j) = P(x_i | A_j)$. Actually, two probabilities are not necessarily equal. Later I found that letting them different as (10)

$$I^*(X; y_j) = \sum_i P(x_i | y_j) \log \frac{P(x_i | A_j)}{P(x_i)} \quad (10)$$

will provide stronger explanatory power.

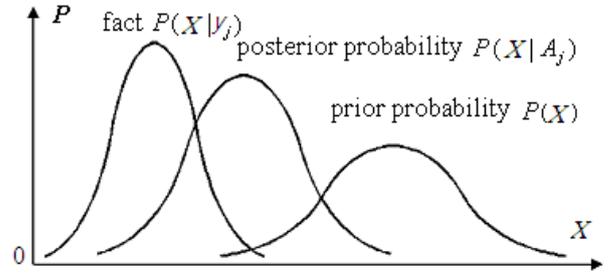

Figure 4.  Illustrating generalized Kullback formula

This is a very important step because this formula indicates that information comes from the test of facts. This is the generalized Kullback formula. In the formula, $P(x_i | y_j)$ means the evidence, $P(x_i | A_j)$ means the forecast according to $y_j$, and $P(x_i)$ means prior knowledge. This formula is similar to Theil's formula for information difference [13], yet it has richer meaning.

We can consider $-\log P(x_i)$ as the coding length of $x_i$ when we code $P(X)$ in the optimal way, consider $-\log P(x_i | A_j)$ as the coding length of $x_i$ when we code $P(X|A_j)$ in the optimal way. Hence $I(X; y_j)$ means the saved average coding length because of $y_j$.

Fig. 4 shows that the more the forecast $P(X|A_j)$ corresponds to the fact $P(X|y_j)$, the more the information is provided, and the maximum information is Kullback information; the farther the prior forecast $P(X)$ is from the fact $P(X|y_j)$ in comparison with $P(X|A_j)$, which means that the forecasted event is more unexpected, the bigger the information amount is.

Now we assume that $X$ is an economical index. We use two parameters $x_j$ and $\sigma_j$ to forecast $X$, the true value function of a prediction $y_j$ is

$$Q(A_j | x_i) = \exp\left[-\frac{(x_i - x_j)^2}{2\sigma_j^2}\right] \quad (11)$$

Then the generalized Kullback formula can be used to evaluate and optimize the predictions about the index.

Let $z$ be some objective factors, $<y_j, \sigma_j>$ be a prediction, the conditional probability of an economical index forecasted be $P(X|z)$. Which prediction $<y_j, \sigma_j>$ is the best? The prediction that makes

$$I^*(X; y_j) = \sum_i P(x_i | z) \log \frac{Q(A_j | x_i)}{Q(A_j)} \quad (12)$$

reach the maximum is the best. The above formula can also be used to select the accuracy $\sigma$ and the pointed position $x_j$ of GPS.

If predictions are often incorrect or predictors lie, the receivers may modify the meanings based on their experiences, such as to let

$$Q(A_j | x_i) = P(y_j | x_i) / \max_k P(y_k | x_i) \text{ for all } x_i \quad (13)$$

so that its maximum is 1. In this way, the information received will increase and be closer to the statistical information.

*E. Generalized Mutual Information Formula*

To calculate the average of $I(X; y_j)$, we get the generalized mutual information formula:

$$I^*(X;Y) = \sum_j \sum_i P(y_j) P(x_i | y_j) \log \frac{P(x_i | A_j)}{P(x_i)} \quad (14)$$

$$= \sum_i P(y_j) P(x_i | y_j) \log \frac{Q(A_j | x_i)}{Q(A_j)}$$

While the forecast corresponds to the fact, $I^*(X;Y)$ becomes

$$I^*(X;Y) = \sum_j \sum_i P(y_j) P(x_i | A_j) \log \frac{Q(A_j | x_i)}{Q(A_j)} \quad (15)$$

In the following discussions about rate distortion and rate tolerance, we can see this kind of formula again.

Now we can see that Shannon mutual information is the special case of the generalized mutual information as the forecast corresponds to the fact; the generalized mutual information also reflects the saved average coding length. Shannon information can be considered as cost and the generalized information can be considered as utility. The former is the upper boundary of the later.

### III. RATE TOLERANCE AND ITS EQUIVALENCES WITH RATE DISTORTION AND COMPLEXITY DISTORTION

*A. From Rate Distortion to Complexity Distortion*

For compressing the data of sound, video, economy, geography, etc., proper lossy coding will substantially improve communication efficiency. For this purpose, Shannon proposed rate distortion theory [2] which was developed by T. Berger [14] and others (see [15], chapter 10).

Assume that we use lossy codes to code a discrete memoryless source $P(X)$ and $X$ is a number. Let $X \in A=\{x_1, x_2 \ldots\}$ for the source and $Y \in B=\{y_1, y_2, \ldots\}$ for the destination, and the distortion between $x_i$ and $y_j$ be denoted by $d_{ij}=d(x_i, y_j)$. Then the average distortion is

$$E(d_{ij}) = \sum_j \sum_i P(x_i, y_j) d_{ij} \quad (16)$$

For given source $P(X)$ and the upper boundary $D$ of $E(d_{ij})$, we change $P(Y|X)$ to seek the minimum $R$ of Shannon mutual information $I(X;Y)$:

$$R(D) = \min_{P(Y|X): E(d_{ij}) \leq D} I(X;Y) \quad (17)$$

$R(D)$ is the rate distortion function. Shannon proved that the minimum average coding length for each number could infinitely approach $R(D)$ [2].

However, in communication practice, we often need to provide tolerance between each pair $(x_i, y_j)$. So, Kolmogorov proposed the structure function [5] based on his complexity theory, Sow proposed complexity distortion [6] also based on Kolmogorev's complexity theory. The complexity theory calls the minimum coding length of a sequence of letters as the complexity of the sequence. For lossy coding, if

$$d(x_i, y_j) = (x_i - y_j)^2 \leq D_c, \text{ for all } i,j \quad (18)$$

is tenable [3], then for each $x_i$, there is a ball ($D$-ball [6]), say $B_i$, on $B$ surrounding $y_i$. It is allowed to use any $y_j \in B_i$ to represent $x_i$. And for each $y_j$, there is a ball, say $A_j$, on $A$ surrounding $x_j$. It is allowed that $y_j$ represents any $x_i \in A_j$. In the complexity theory, all balls have the same radius [4] $D_c^{0.5}$. For given $P(X)$ and $D_c$, we may get the minimum average coding length (block-coding might be required) or the minimum Shannon mutual information (their tiny difference is neglected here) $C$, which is the function of $D_c$, i. e. $C=C(D_c)$, which is the complexity distortion function. When using the same coding scheme, we assume that each $A_j$ contains the same number of elements and the number is $S=|A_j|$, then $h_x(D_c) = \log S$ is called the structure function.

However, we also need different size tolerance sets or balls to limit the coding errors. For example, with the brightness of colors increasing, the sensitivity of human eyes to color differences will decrease, which means that we need smaller tolerance balls for darker colors and bigger tolerance balls for brighter colors.

*B. Defining Rate Tolerance and Proving that Complexity Distortion is Its Special Case*

Assume there is a similarity relation on AXB. The similarity degree between $x_i$ and $y_j$ is $c_{ij}=c(x_i, y_j)=c(x_i, x_j) \in [0,1]$, such as defined by (2). For each $x_i$, there is a $B_i$ and the membership grade of $y_j$ in $B_i$ is $c_{ij}$. For each $y_j$, there is a $A_j$ on $A$ and the membership grade of $x_i$ in $A_j$ is also $c_{ij}$. That means that $B_i$ and $A_j$ are fuzzy sets and

$$Q(B_i | y_j) = Q(A_j | x_i) = c_{ij} \quad (19)$$

If for each $B_i$, there is the constraint

$$P(y_j | x_i) \leq P(y_j | B_i), \text{ for } y_j \in B \text{ as } Q(A_j|x_i) \neq 1 \quad (20)$$

We call $T=\{B_1, B_2 \ldots\}$ as the tolerance of source $P(X)$. If $B_1$, $B_2 \ldots$ are all clear sets, i.e. $c_{ij} \in \{0,1\}$. The above constraint becomes

$$\sum_j P(y_j | x_i) = 0, \text{ for all } y_j \notin B_i \quad (21)$$

Now, we select $y_j$ to represent $x_i$. For given $P(X)$ and $T$, we change $P(Y|X)$ to seek the minimum Shannon mutual information $R$. That is

$$R(T) = \min_{P(Y|X): P(Y|X) \leq P(Y|T) as Q(A_j|x_i) \neq 1} I(X;Y) \quad (22)$$

$R(T)$ is the function of rate tolerance. Obviously, if $B_1$, $B_2$, … are the same size balls or regions which are clear sets, then $R(T)$ will become $C(D_c)$. So, $C(D_c)$ is a special case of $R(T)$.

---

[3] I use $D_c$ here instead of $D$ because $D$ means average distortion.
[4] Generally, block coding is used to explain complexity distortion, and the radius is of $n$-dimension space.

In the following, we assume $B_1, B_2 \ldots$ are clear sets, and then examine the equivalence between $R(D)$ and $R(T)$.

Now consider function $R(D)$. Let us define the distortion of $y_j$ relative to $x_i$:

$$d_{ij} = d(x_i, y_j) = -\log c_{ij} = \begin{cases} 0, & c_{ij} = 1, \\ \infty, & c_{ij} = 0. \end{cases} \text{ for all } i,j \quad (23)$$

According to formula (21) and (23), the constraint $E(d_{ij}) \leq D=0$ is equivalent to the constraint $T$, i. e. $R(D=0)=R(T)$. So, $C(D_c)$ is also a special case of $R(D)$.

C. *A Generalized Entropy $H^*(X)$ Reflecting Rate Distortion and Complexity Distortion*

Now we prove that $R(D=0)$ equals a generalized entropy while $T$ is a group of clear sets.

We know the parameter form [14] of

$$R(D): \begin{cases} D(s) = \sum_i P(x_i) P(y_j) \exp(sd_{ij}) \lambda_i d_{ij} \\ R(s) = sD(s) + \sum_i P(x_i) \log \lambda_i \end{cases} \quad (24)$$

where

$$\lambda_i = 1 / \sum_j P(y_j) \exp(sd_{ij}) \quad (25)$$

Let $s = -1/(2\sigma^2)$, we can see clear relationship between GPS information and rate distortion function. We can consider $\exp(sd_{ij}) = Q(B_i|y_j)$ and $\lambda_i = 1/Q(B_i)$. Then we have

$$R(s) = \sum_i \sum_j P(x_i) P(y_j | B_i) \log \frac{Q(B_i | y_j)}{Q(B_i)} \quad (26)$$

which is similar to (15). When $T$ is a group of clear sets, $\exp(-\infty)=0$, $\exp(0)=1$, $D=0$, $R$ is irrelative to $s$. Hence

$$R(s) = H^*(X) = -\sum_i P(x_i) \log Q(B_i) \quad (27)$$

where $Q(B_i)$ is the function of $P(Y)$. It is not true that any $H^*(X)$ equals a $R(D=0)$. Only the minimum of $H^*(X)$ obtained by changing $P(Y)$ is $R(D=0)$. That is

$$R(D=0) = H_{P(Y)}^*(X) = \min_{P(Y)}[-\sum_i P(x_i) \log Q(B_i)] \quad (28)$$

So, $R(T) = R(D=0) = H_{P(Y)}^*(X)$. Only when all $B_i$ have the same size, is there $C(D_c) = R(T) = R(D=0)$. The conditional entropy of $X$ with $T$ as the condition is

$$H(X | T) = \sum_j \sum_i P(x_i) P(x_i | A_j) \log P(x_i | A_j) \quad (29)$$

If all $A_j$ are the same size balls and $|A_j|=S$ and $P(X|A_j)$ are equiprobable for all $j$, then $H(X|T)$ will become the structure function $h_x(D_c)$.

Obviously, it is not correct to conclude $C(D_c)=R(D_c)$ [6] in most situations or to confirm that there is a general equivalence between the distortion-rate and the expected structure function [5], because in the complexity theory, those constraint sets are clear, yet in the rate distortion theory, those constraint sets are fuzzy. The later has looser constraints than the former. We use a coding example to explain their differences.

D. *Using a Coding Example to Explain $R(D_c)<C(D_c)$*

Now we prove that $R(D=0)$ equals a generalized entropy while $T$ is a group of clear sets.

This example will provide the thread for changing $P(Y|X)$ to decrease $I(X;Y)$.

Example: $A=\{1, 2, 3, 4\}$ and $B=(1, 2, 3, 4)$. $P(X)$ is equiprobable. The coding is to use $y_j$ in $B$ to represent $x_i$ in $A$ with tolerance $|X-Y|\leq 1$。

Real procedure for data compression is $P(X) \rightarrow$ source coding $\rightarrow$ storing and transmitting $\rightarrow$ Decoding $\rightarrow P(Y)$. However we only consider what $P(Y|X)$ can result in the minimum $I(X;Y)$ without considering middle details where block-coding might be required.

The equation (28) reminds us that for each $x_i$, it is better when $Q(B_i)$ is bigger; also it is better when different tolerance sets overlap each other more. If so, $H(X|Y)$ will bigger, $I(X;Y) = H(X)-H(X|Y)$ will be less.

If we simply use $y_2$ to represent $x_1$ and $x_2$, and use $y_3$ to represent $x_3$ an $x_4$, then the average information is 1 bit. Now we refer to (25) and (28) to code the source as follows: when $X=1$, let $Y=2$; when $X=4$, let $Y=3$; when $X=2$ or 2, let $Y=2$ or 3 randomly. See Table 1 for some detailed data.

Note $Q(B_2)=Q(B_3)=1$, so $X=2$ or 3, the information amount is 0. Hence $R(T)=C(D_c)=-0.25\log 0.5 - 0.25\log 0.5 = 0.5$ bit. However in this way, $E(d_{ij})=0.75<1$. The following coding can prove that $R(D_c)$ may be less then $C(D_c)$.

Now let $d_{ij}=(y_j-x_i)^2$ and distortion $E(d_{ij}) \leq D=1$ to seek $R$. Let $s=-0.45$, $y_2=y_3=0.5$, $Q(A_i|y_j)=\exp[s(y_j-x_i)^2]$, $P(y_j|x_i)=P(y_j|B_i)$. Hence, we get the data shown in Table 2.

Then according to (24), $D=0.994$, $R(D)=sD+H^*(X)=-0.447+0.816=0.369$ bit$<0.5$ bit. It can be seen that generally $R(D_c)<C(D_c)$.

TABLE I. SEEK $P(Y|X)$ FOR $C(D_c=1)$

| $P(x_i)$ | $x_i$ | $y_j$ | $P(y_j)$ | $Q(B_i)$ | $-P(x_i)\log Q(B_i)$ |
|---|---|---|---|---|---|
| 0.25 | 1 | 1 | 0 | $P(y_2)=0.5$ | 0.25 |
| 0.25 | 2 | 2 | 0.5 | $P(y_2)+P(y_3)=1$ | 0 |
| 0.25 | 3 | 3 | 0.5 | $P(y_2)+P(y_3)=1$ | 0 |
| 0.25 | 4 | 4 | 0 | $P(y_4)=0.5$ | 0.25 |

TABLE II. SEEK $P(Y|X)$ FOR $R(D=1)$

| $P(x_i)$ | $x_i$ | $P(y_2|x_i)$ | $P(y_3|x_i)$ | $P(y_j)$ | $Q(B_i)$ | $-P(x_i)\log Q(B_i)$ |
|---|---|---|---|---|---|---|
| 0.25 | 1 | 0.8 | 0.2 | 0 | 0.395 | 0.335 |
| 0.25 | 2 | 0.903 | 0.097 | 0.5 | 0.816 | 0.073 |
| 0.25 | 3 | 0.097 | 0.903 | 0.5 | 0.816 | 0.073 |
| 0.25 | 4 | 0.2 | 0.8 | 0 | 0.395 | 0.335 |

## E. Rate Tolerance and Its Equivalence to Rate Distortion

Let $T=\{B_1, B_2 \ldots\}$ be a group of fuzzy sets. The generalized mutual information is $I(X;Y)= H(Y) - H(Y|X)$. The tolerance is $T$ or inequality (20). It is easy to prove that for given $P(Y)$, $H(Y|X)$ reaches it's the maximum while

$$P(y_j | x_i) = P(y_j | B_i), \text{ for all } i,j \qquad (30)$$

This equation is necessary but sufficient for $R$. To change $P(Y)$ to make $I(X;Y)$ reach the minimum, we have the rate tolerance function:

$$R(T) = \min_{P(Y)} \sum_i \sum_j P(x_i) P(y_j | B_i) \log \frac{Q(B_i | y_j)}{Q(B_i)} \qquad (31)$$

Assume $Q(B_i|y_j)=\exp(sd_{ij})$ for all $i, j$ and refer to Section III-C, we have $R(T)= R(D)$. That means any $R(D)$ is a special case of $R(T)$ as $Q(B_i |y_j)=\exp(sd_{ij})$ for all $i,j$. The $s$ reflects the accuracy of a prediction. The bigger it's absolute value is, the bigger the $R$ is.

There exists obvious relationship between $R(D)$ and the generalized information measure.

## IV. SUMMARY

In this paper I use GPS as an example to derive the generalized information formulae from the classical information formulae, discuss how rate tolerance and rate distortion are related to the generalized mutual information, prove that rate distortion $R(D)$ is a special case of rate tolerance $R(T)$ and that complexity distortion $C(D_c)$ is a special case of $R(D)$. In my previous researches [1] [4] I also studied the relationship between the maximum entropy principle in thermodynamics and rate tolerance, and proposed that by replacing distortion $D$ with the lower boundary $G$ of generalized mutual information, we can get the function $R(G)$ where $G/R$ indicates communication efficiency including the efficiency in which the sender's lies result in the information loss of its enemy. I will discuss these topics further elsewhere.

## REFERENCES

[1] C. Lu, A Generalized Information Theory (in Chinese), China Science and Technology University Press , 1993.

[2] C. E. Shannon, "A mathematical theory of communication", Bell System Technical Journal, Vol.27, pp. 379-429 and 623-656, 1948.

[3] C. Lu, "Meanings of generalized entropy and generalized mutual information for coding" (in Chinese), J. of China Institute of Communication, vol. 15, No.6, pp. 37-44, 1994..

[4] C. Lu, "A generalization of Shannon's information theory", Int. J. of General Systems, vol. 28, No. 6, pp. 453-490, 1999.

[5] P. D. Grunwald and P. B. Vitany, "Kolmogorov Complexity and Information Theory With an Interpretation in Terms of Questions and Answers", Journal of Logic, Language and Information, pp. 497-529, Dec. 2003.

[6] D. M. Sow, "Complexity Distortion Theory", IEEE Transactions on Information Theory, vol. 49, No. 3, pp. 604-609, March 2003.

[7] P. Z. Wang, "Random sets in Fuzzy Set theory", In: System & Control Encyclopedia, edited by M. G. Singh, Pergamon Press, pp. 3945-3947, 1987.

[8] J. B. Tenenbaum and T. L. Griffiths, "Generalization, similarity, and Bayesian inference", Behavioral and Brain Science, vol. 24, No.4, pp. 629-640, 2001.

[9] L. A. Zadeh, "Fuzzy sets", Infor. Contr., vol. 8, pp. 338-353, 1965.

[10] L. A. Zedah, "Probability measures of fuzzy events", Journal of mathematical Analyses and Applications. vol.23, pp. 421-427, 1968..

[11] S. Kullback, Information and Statistics, John Wiley & Sons Inc., New York,1959.

[12] [12] K. R. Popper, Conjectures and Refutations: The Growth of Scientific Knowledge, Harper & Row, Publishers, New York and Evanston, 1968.

[13] [13] H. Theil, Economics and Information Theory, North-Holland, Amsterdam, 1967.

[14] T. Berger, Rate Distortion Theory: A Mathematical Basis for Data Compression, Englewood Cliffs, NJ: Prentice-Hall, 1971.

[15] T. M. Cover and J. A. Thomas, Elements of Information Theory, Second Edition, John Wiley & Sons, 2006.